\documentclass[a4paper,11pt]{article}

\let\latexput\put
\usepackage{pictexwd}

\pdfoutput=1 

\usepackage{jheppub} 
\usepackage{amsmath}                    
\usepackage[bbgreekl]{mathbbol}

\usepackage{amsfonts,amssymb,amsthm,mathtools} 
\usepackage[T1]{fontenc} 

\usepackage{psfrag}
\usepackage{indentfirst} 
\usepackage{bm}          
\usepackage[usenames,dvipsnames]{xcolor}
\usepackage{braket}
\usepackage{bbm}
\usepackage{dsfont}
\usepackage{amssymb,amsfonts,latexsym,cancel}
\usepackage{ulem}
\usepackage{amsthm}    
\usepackage{wrapfig}
\usepackage{subfigure} 
\usepackage{rawfonts}
\usepackage{pictexwd}
\usepackage{soul}
\usepackage{url}
\usepackage{enumitem}
\usepackage{MnSymbol}
\usepackage{hyperref}
\usepackage{tensor}
\usepackage[titletoc,title]{appendix}
\usepackage{mathtools}
\usepackage{xcolor} 
\usepackage{scalerel}
\usepackage{graphicx}
\usepackage{todonotes}
\usepackage{upgreek}

\DeclareSymbolFontAlphabet{\mathbbm}{bbold}
\DeclareSymbolFontAlphabet{\mathbb}{AMSb}
\DeclareSymbolFontAlphabet{\mathbbl}{bbold}

\AtBeginDocument{}%

\makeatletter
\newcommand\rrule[3][0pt]{%
	\ifdim#2>#3\math@hrule[#1]{#2}{#3}\else\math@vrule[#1]{#2}{#3}\fi}
\newcommand\math@hrule[3][0pt]{%
	\gdef\mystery@factor{0.07}%
	\@tempdima=#3%
	\rule[#1]{0pt}{#3}
	\raisebox{.5\@tempdima+#1}{%
		\makebox[#2][l]{\kern-.5\@tempdima\@@mathrule{#2}{#3}}}%
}
\newcommand\math@vrule[3][0pt]{%
	\gdef\mystery@factor{0.0}%
	\@tempdima=#2%
	\rule[#1]{0pt}{#3}
	\raisebox{-.0\@tempdima+#1}{%
		\kern0.5\@tempdima%
		\rotatebox{90}{\kern-0.5\@tempdima\makebox[#3][l]{\@@mathrule{#3}{#2}}}%
		\kern0.5\@tempdima}%
}
\def\@@mathrule#1#2{%
	\@tempdimb=#2%
	\@tempdima=\dimexpr#1-\mystery@factor\@tempdimb
	\pdfliteral{%
		q []0 d %
		1 J 
		\strip@pt\@tempdimb\space w \strip@pt\@tempdimb\space 0 m %
		\strip@pt\@tempdima\space 0 l S Q }}
\makeatother

\makeatletter
\DeclareFontFamily{OMX}{MnSymbolE}{}
\DeclareSymbolFont{MnLargeSymbols}{OMX}{MnSymbolE}{m}{n}
\SetSymbolFont{MnLargeSymbols}{bold}{OMX}{MnSymbolE}{b}{n}
\DeclareFontShape{OMX}{MnSymbolE}{m}{n}{
<-6>  MnSymbolE5
<6-7>  MnSymbolE6
<7-8>  MnSymbolE7
<8-9>  MnSymbolE8
<9-10> MnSymbolE9
<10-12> MnSymbolE10
<12->   MnSymbolE12
}{}
\DeclareFontShape{OMX}{MnSymbolE}{b}{n}{
<-6>  MnSymbolE-Bold5
<6-7>  MnSymbolE-Bold6
<7-8>  MnSymbolE-Bold7
<8-9>  MnSymbolE-Bold8
<9-10> MnSymbolE-Bold9
<10-12> MnSymbolE-Bold10
<12->   MnSymbolE-Bold12
}{}

\let\llangle\@undefined
\let\rrangle\@undefined
\DeclareMathDelimiter{\llangle}{\mathopen}%
{MnLargeSymbols}{'164}{MnLargeSymbols}{'164}
\DeclareMathDelimiter{\rrangle}{\mathclose}%
{MnLargeSymbols}{'171}{MnLargeSymbols}{'171}
\makeatother

\def\wwedgee{{\setbox0\hbox{\ensuremath{\mathrel{\wedge}}}\rlap{\hbox to \wd0{\hss\hspace*{.6ex}\ensuremath\wedge\hss}}\box0}}

\usepackage{float}

\newcommand{\avg}[1]{\left\langle{#1}\right\rangle}

\title{\boldmath Structure of the medium formed in heavy ion collisions}

\author[a]{J. R. Alvarado Garc\'ia,}
\author[a]{D. Rosales Herrera,}
\author[a]{A. Fernández Téllez,} 
\author[b,c, d]{Bogar D\'{\i}az,}
\author[e]{and J. E. Ram\'irez}

\affiliation[a]{Facultad de Ciencias F\'isico Matem\'aticas, Benem\'erita Universidad Aut\'onoma de Puebla,
Apartado Postal 165, 72000 Puebla, Puebla, Mexico}
\affiliation[b]{Departamento de Matem\'aticas, Universidad Carlos III de Madrid, Avenida  de la Universidad 30, 28911 Legan\'es, Spain}
\affiliation[c]{Grupo de Teor\'ias de Campos y F\'isica Estad\'istica. Instituto Gregorio Mill\'an (UC3M), Unidad Asociada al Instituto de Estructura de la Materia, CSIC, Serrano 123, 28006 Madrid, Spain}
\affiliation[d]{Instituto de Ciencias Matem\'aticas (ICMAT), CSIC-UAM-UC3M-UCM, C. Nicol\'as Cabrera 13-15, 28049 Madrid, Spain}
\affiliation[e]{Centro de Agroecología, Instituto de Ciencias, Benemérita Universidad Autónoma de Puebla, Apartado Postal 165, 72000 Puebla, Puebla, Mexico}

\emailAdd{bodiazj@math.uc3m.es}
\emailAdd{jhony.ramirezcancino@viep.com.mx}

\abstract{We investigate the structure of the medium formed in heavy ion collisions using three different models: the Color String Percolation Model (CSPM), the Core-Shell-Color String Percolation Model (CSCSPM), and the Color Glass Condensate (CGC) framework. We analyze the radial distribution function of the transverse representation of color flux tubes in each model to determine the medium's structure. Our results indicate that the CSPM behaves as an ideal gas, while the CSCSPM exhibits a structural phase transition from a gas-like to a liquid-like structure. Additionally, our analysis of the CGC framework suggests that it produces systems that behave like interacting gases for AuAu central collisions at RHIC energies and liquid-like structures for PbPb central collisions at LHC energies.}

\arxivnumber{}

\let\put\latexput

\begin{document}
\maketitle
\flushbottom

\section{Introduction}\label{sec_introduction}

Quark-Gluon Plasma (QGP) is a state of matter that is believed to have existed in the early universe, shortly after the Big Bang. In this state, the quarks and gluons (which are normally confined within protons and neutrons) are liberated from their confinement and move freely in a hot and dense soup. The existence of QGP was first proposed in the 1970s \cite{SHURYAK1978, McLerran1981, Bjorken1983} trying to understand the behavior of high-energy collisions between atomic nuclei.

It was not until the late 1990s, however, that experiments provided evidence for the creation of QGP. The collision of heavy ions, such as Au-Au at RHIC, produced a \textit{liquid} made of quarks and gluons (the QGP), which exhibited a lower ratio of shear viscosity over entropy density than any other known material  \cite{gyulassy, ADAMS2005102, ADCOX2005184}. Its existence was later confirmed in Pb-Pb collisions at LHC \cite{PhysRevLett.105.252302,2012330, CMS}. Also, similar properties have been observed in other collisions, such as pp, pA collisions at LHC \cite{CMS2}, as well as d-Au and $^3$He-Au at RHIC \cite{Aidala2019}.

Several models have been proposed in the literature to describe the quark-gluon plasma formation and its phenomenology. They are able to explain most of the experimental data, but not all. Two of the most popular approaches are the Color String Percolation Model (CSPM) \cite{BRAUN20151, string, Braun2000, PRLPajares} and the Color Glass Condensate model (CGC) \cite{Tribedy:2010ab,Tribedy:2011aa}.

We are interested in studying the capacity of the two mentioned models to predict the liquid behavior of the QGP. To be precise, we analyze three models a) CSPM, b) the Core-Shell-Color String Percolation Model (CSCSPM) \cite{JBP2021}, which is a generalization of the CSPM, and c) CGC. One way to analyze the physical structure of a system is by examining the behavior of the radial distribution function (also referred to as the pair correlation function), denoted as $g(r)$. It explains the variation of the average particle count concerning the distance $r$ from a given point. This function is a prevalent tool for evaluating packing structures and provides valuable data on the long-range correlations between particles and their organization \cite{ASTE200632}. The structure of the systems is revealed by $g(r)$ as follows: a) if it is featureless (flat function) the system corresponds to an ideal gas structure, b) for a non-ideal gas the $g(r)$ possess a peak (a maximum) follows for a plain behavior, and c) for a liquid-like structure $g(r)$ exhibits several peaks (increasingly smaller).

In this work, we review the radial distributions function for the CSPM and the CSCSPM that was studied in \cite{JBP2021}, and analyze it for the CGC model. We present the basic features of the three mentioned models in section \ref{models}. Then, in section \ref{simulations} we discuss the employed simulation methods and we analyze the behavior of $g(r)$ for the models of section \ref{models}. Finally, we give our conclusions in section \ref{conclu}.

\section{Models}\label{models}

In this section, we present the general features of the CSPM, CSCSPM, and CGC models.

\subsection{Color string percolation model}

One of the main models used to describe the formation of quark-gluon plasma is the CSPM. It has successfully explained many of the experimental findings on collisions involving pp, pA, and AA interactions, including the azimuthal distributions of the resulting particles and the temperature dependence of the ratio between the shear and bulk viscosities over the entropy density \cite{P2016, Sahoothermo}. The CSPM is also able to predict the center of mass energy required for the QGP formation depending on the size of the systems \cite{letter}.

The CSPM characterizes ion collision by the strong interaction between their constituent partons, which can be represented by color flux tubes extended in the beam axis direction, depicted as circular areas in the transverse plane. These objects, named strings, are represented by fully penetrable disks of radius $r \sim$ 0.2-0.3 fm \cite{Amelin:1993cs,Braun2000,BRAUN20151,string}. The systems under this picture can be studied from percolation theory, where the overlapping between strings gives rise to different regions called color sources. The number of strings grows with the system's size, multiplicity, and energy, allowing the formation of a spanning cluster that marks the geometric phase transition and then the onset of the QGP formation.

Each individual color string has a certain area $S_1$, color charge arbitrarily oriented $\mathbf{Q}_1$ with its corresponding color field intensity $Q_1$, and produces particles in the midrapidity region with multiplicity $\mu_1$, and average squared transverse momentum $\langle p_T ^2 \rangle_1$ \cite{Braun2000,BRAUN20151,string}.
Because of the random orientation of the color field, the resulting intensity of the color field for a color source, indexed by $i$, is proportional to the square root of the number of overlapped string $n_i$ in the fraction of the string area $S^{(i)}/S_1$. Therefore, the obtained color field intensity is $Q^{(i)} = \sqrt{n_i}Q_1 S^{(i)}/S_1$ \cite{Braun2000,BRAUN20151,string}.

As a consequence of the color sum, the multiplicity and average squared transverse momentum of a color source is obtained via $\mu^{(i)} =\mu_1 Q^{(i)}/Q_1 = \sqrt{n_i}\mu_1 S^{(i)}/S_1 $ and $\avg{p_T}^{(i)} = \avg{p_T^2}_1 Q^{(i)}/Q_1 = \sqrt{n_i}\avg{p_T^2}_1 S^{(i)}/S_1 $. The sum of all contributions to a cluster of $M$ color sources and $n$ strings produces the total $\mu_n$ and $\avg{p_T^2}_n$ of the cluster \cite{Braun2000,BRAUN20151}.
Let us consider the two extreme cases of a cluster of $n$ strings: on one hand, for the case of strings just touching each other, the total area covered by disks is the sum of the individual areas $S_n = nS_1$, and the same for the total multiplicity given by the sum of individual contributions $\mu_n = n\mu_1$, the average of $p_T^2$ match with the weighted sum of each string $ \langle p_T ^2 \rangle_n = n\mu_1\avg{p_T^2}/\mu_n = \langle p_T ^2 \rangle_1$. On the other hand, the case of $n$ strings fully overlapped leads to a total area $S_n = S_1$, and the only color source has the same area $S_1$; the observables $\mu_n = \sqrt{n}\mu_1$ and $\avg{p_T^2}_n = \sqrt{n} \avg{p_T^2}_n$ correspond to the multiplicity and average of $p_T^2$ of a single color source with $n$ overlaps, in this scenario it is found the maximal suppression for the multiplicity and the average of $p_T^2$ is maximally enhanced. This is consistent with the conservation of momentum relationship $n \mu_1 \avg{p_T^2}_1 = \mu_n \avg{p_T^2}_n$, that arises from the color sum rules established in the model \cite{Braun2000,BRAUN20151}.

Note that a suppression effect naturally emerges depending on the spatial distribution of the strings, this effect is taken into account by the color reduction factor, $F(\eta)$, which depends on the fraction of the total area $S$ occupied by $n$ strings, the so-called string density parameter $\eta = n S_1/S$ \cite{Braun2000,BRAUN20151}. The color reduction factor drives the suppression of the multiplicity ($\mu = nF(\eta)\mu_1$) and the enhancement of the transverse momentum ($\avg{p_T^2} =\avg{p_T^2}_1 /F(\eta) $) \cite{DIASDEDEUS2006455}. 
It was found that $F(\eta)$ is a decreasing function, which means that as the string density increases, a colliding system produces fewer charged particles per string but with higher momentum.
In this way, the color suppression factor emerges from the color string clustering process \cite{letter}.


\subsection{Core-shell-color string percolation model}

In \cite{JBP2021} was proposed a hybrid core-shell model together with the traditional CSPM, called CSCSPM. The idea was to take into account an excluding or repulsive interaction between strings. To be precise, the authors introduced a concentric region of exclusion  into the strings (core region, the rest of the string area is called the shell region) of diameter $\lambda \sigma$ ($0\leq\lambda\leq 1$, $\sigma$ denotes the diameter of the strings). They also introduced a probability $q_\lambda$ that determines whether a string can overlap with another string in its core region, classifying strings as soft or hard depending on whether they permit such overlaps. It is important to note that this overlap condition applies solely to core-core interactions, as core-shell and shell-shell overlaps are permitted. 

In the same way that their predecessor, the CSCSPM model explains particle production in collision physics through the formation of string clusters (same strings as in the CSPM). Notice that the hard strings act as a fluid of hard disks of diameter $\lambda \sigma$, while the soft ones are still ideal particles. Moreover, the parameter $q_\lambda$ modules the number of hard strings distributed in the system.
However, the structures and phenomenology of the system depend on the combinations of both parameters.
For example, if $\lambda=1$ and $q_\lambda=0$, the system recovers the picture of a fluid of hard disks \cite{chandler}, which may exhibit a liquid structure for densities above a particular string density. This condition prevents the formation of clusters, and then each string should produce charged particles individually, but on average, their transverse momentum squared will be the corresponding of one string divided by the multiplicity.
This mechanism inhibits the formation of charged particles with higher transverse momentum. Also, if $q_\lambda=0$, the model reproduces the continuum percolation of disks with hard cores \cite{hcperc,hcperc2}. Finally,
If $\lambda=0$ or $q_\lambda=1$, the system corresponds to traditional 2D continuum percolation \cite{continuumperc,mertens}, which is the geometric picture of the CSPM.

It was found in Ref.~\cite{JBP2021} that combinations of parameter values exist such that, at the same time, allow the system the clustering of color strings and the formation of coordination shells, the main indications of the changes in the structure of the system.

\subsection{Color glass condensate model}

The CGC can provide the initial conditions needed to estimate the transport coefficients and bulk properties of the strong-interacting matter created in heavy ion collisions \cite{Albacete:2014fwa}. It is characterized by strong coherent gluon fields leading to parton saturation controlled by a dynamically generated transverse momentum scale, the saturation scale $Q_s$ \cite{Gribov:1983ivg, Mueller:1989st, Blaizot:1987nc, McLerran:1994vd, Balitsky:1995ub, Kovchegov:1999yj, Jalilian-Marian:1997jhx, Jalilian-Marian:1997qno, Iancu:2001ad, Iancu:2000hn, Weigert:2000gi}.  
The high gluon densities correspond to strong classical fields, and the quantum corrections to them are incorporated via nonlinear renormalization group equations. The nonlinear density-dependent terms in the CGC evolution equations can be identified as gluon recombination processes that saturate the increment in gluon densities below $Q_s$. Then, the value of $Q_s$ sets the scale of gluon field fluctuations that can describe the bulk multiplicity and its fluctuations depending on the considered colliding system \cite{Dumitru:2011wq, Dumitru:2012yr}. 
The particle multiplicity distributions are intrinsically dependent on event-by-event fluctuations of the incoming nuclear wave functions. 

In the framework of the CGC, to provide information about the systems' characteristics it is used the Impact Parameter (IP) Glasma model, which  combines the IP saturation dipole model \cite{Kowalski:2003hm} and nonlinear dynamics of the gluon fields \cite{Schenke:2012wb,Schenke:2012hg}. The $Q_s^2$'s distribution can be used to sample different color field configurations inside the proton according to the McLerran-Venugopalan (MV) model \cite{McLerran:1993ni,McLerran:1993ka}. The IP-Glasma model includes fluctuations in the nuclear and sub-nuclear position of static large-$x$ color charges leading the gluon fields fluctuations that describe the dynamical small $x$ modes in the CGC effective field theory \cite{Gelis:2010nm, Schenke:2013dpa}. The different occupation probabilities of the QCD fields give rise to global fluctuations in observables such as the energy density in the transverse plane of the collision, which lead to fluctuations in the charged hadron flow harmonics through the dynamical evolution of the system \cite{Gale:2012rq}.
It was proposed that in every event, the value of $Q_s^2$ at every point in the transverse plane fluctuates according to a probability distribution given by \cite{Dusling:2012iga, McLerran:2015qxa}
\begin{equation}
P\left[ \ln\left(\frac{Q_s^2}{\avg{Q_s^2}} \right) \right] = \frac{1}{\sqrt{2\pi \varsigma^2}} \exp\left[ - \frac{1}{2 \varsigma^2} \ln^2\left(\frac{Q_s^2}{\avg{Q_s^2}} \right)\right].
\label{eq.PQs}    
\end{equation}
This particular distribution gives rise to a skewed distribution of $Q_s/\avg{Q_s}$ around 1. We use the value of $\varsigma = 0.6$ reported in \cite{Schenke:2020mbo} to describe heavy ion collisions from AuAu to PbPb. 

In addition to the skewness, the saturation scale $Q_s$ also depends on the size and energy of the colliding nuclei. It was found that $Q_s$ grows with the size of the colliding nuclei as $\sim A^{1/3}$ \cite{McLerran:1993ni, McLerran:1993ka, Kovchegov:1996ty}, where $A$ is the nucleon number. Moreover, $Q_s$ rises with the energy as the power law $\sim \sqrt{s}^\lambda$ (at midrapidity region) with $\lambda = 0.252$ \cite{Kharzeev:2004if}. For AuAu collisions at $\sqrt{s} =$ 130 GeV, the saturation scale is taken to be $\avg{Q_s^2} =$ 2 GeV$^2$ \cite{Kharzeev:2000ph}. Using the scaling laws described above, the saturation scale for AuAu central collisions at $\sqrt{s} =$ 200 GeV and PbPb central collisions at $\sqrt{s} =$ 2760 and 5020 GeV are estimated to be 2.23, 4.399, and 5.114 GeV$^2$, respectively. These values are used in the subsequent analysis of the heavy ion collisions.

\section{Simulation methods and data analysis}\label{simulations}
In this section, we discuss the methodology used to determine the radial distribution functions and provide an introduction to the computer simulation basic notions of the models of interest. We also present the obtained results.

The radial distribution function $g(r)$ (also referred to as a pair correlation function or pair distribution function) determines the structure of the matter by analyzing the variations in the positions of its constituents, more precisely, in the average of the local density number at a distance $r$. Specifically, $g(r)$ describes the average local density of objects at a distance $r$ from a reference object. In our simulations, we generate samples of the systems as square boxes of size $L=8\sigma_\circ$, with $\sigma_\circ$ being the characteristic diameter of the objects distributed on the transverse plane of each model.
Next, we generate suitable configurations of the system that satisfy the conditions of the corresponding model.
Then, the radial distribution function is estimated as follows
\begin{equation}
    g(r)=\frac{n(r)L^2}{N(2\pi \Delta r (r+0.5\Delta r)+\pi\Delta r^2)},
\label{eq:gr}
\end{equation}
where $n(r)$ is the average number of objects at a distance between $r$ and $r+\Delta r$ from a trial disk allocated on the center of the square box, and $N$ is the number of objects distributed on the square box.

In the following subsections, we deeply discuss the peculiarities of the simulation and structures observed for the CSPM, CSCSPM, and CGC models. In order to compare the results of the different models, we compute $g(r)$ over the range of distances $r/\sigma_\circ$ from 0 to 3.5, with increments of $\Delta r/\sigma_\circ=0.035$.

\subsection{Color string percolation model}

\begin{figure}[H]
\centering
\includegraphics[scale=0.5]{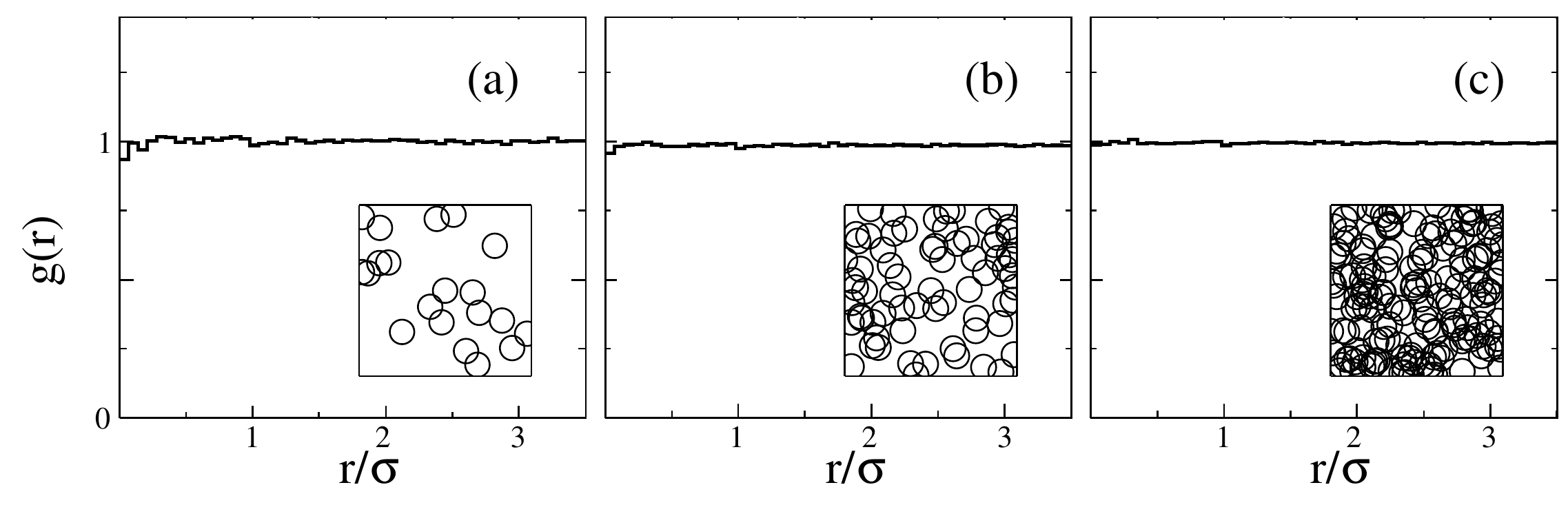}
\caption{Radial distribution function for CSPM systems.}
\label{fig:cspm-gr}
\end{figure}

As we explain in section \ref{models}, the CSPM considers strings like fully penetrable disks. This is equivalent to the image of a classical ideal gas and has to be reflected in the radial distribution.  
In this model, the simulations are performed accordingly to the two dimensional continuum percolation of disks. In this way, the color strings are uniform and randomly allocated in the square box without any conditions on the overlapping between them.
After adding $N$ disks, the radial distribution function is computed by using \eqref{eq:gr}. This was previously analyzed in \cite{JBP2021}, and it was found that independently of the string density the $g(r)$ is a flat function. We show in Fig.~\ref{fig:cspm-gr} our results of $g(r)$ for the CSPM considering different string densities.

We must mention that other authors have also predicted the ideal gas behavior of the CSPM. For instance, in Refs.~\cite{sahoo, sahoo2} the authors studied the electrical and thermal conductivity of the QGP, using the CSPM, founding that the system agrees with an ideal gas. Also, in Ref.~\cite{jerc}, a finite-size analysis of the speed of sound was performed, concluding that the CSPM corresponds to a mean-field theory.

\subsection{Core-shell-color string percolation model}

As discussed above, the CSPM only describes systems with an ideal gas structure because of the fully penetrable color strings. However, other structures (as nonideal gas or liquid-like) may be observed if a repulsive interaction between the strings is introduced, as was done in the CSCSPM.

The CSCSPM configurations are generated by using the random sequential addition algorithm, in which disks are added one by one. At each step, a test string is randomly placed on the square box and designed as soft or hard. Then, according to the values of $\lambda$ and $q_\lambda$, the test string is accepted if it, along with its neighbor strings, satisfies a valid configuration.
Otherwise, the test string is rejected and a new test string is checked.
This procedure is repeated until the system is filled with $N$ strings.
After generating the CSCSPM configuration, we measure its g(r) using \eqref{eq:gr}.

\begin{figure}[H]
\centering
\includegraphics[scale=0.5]{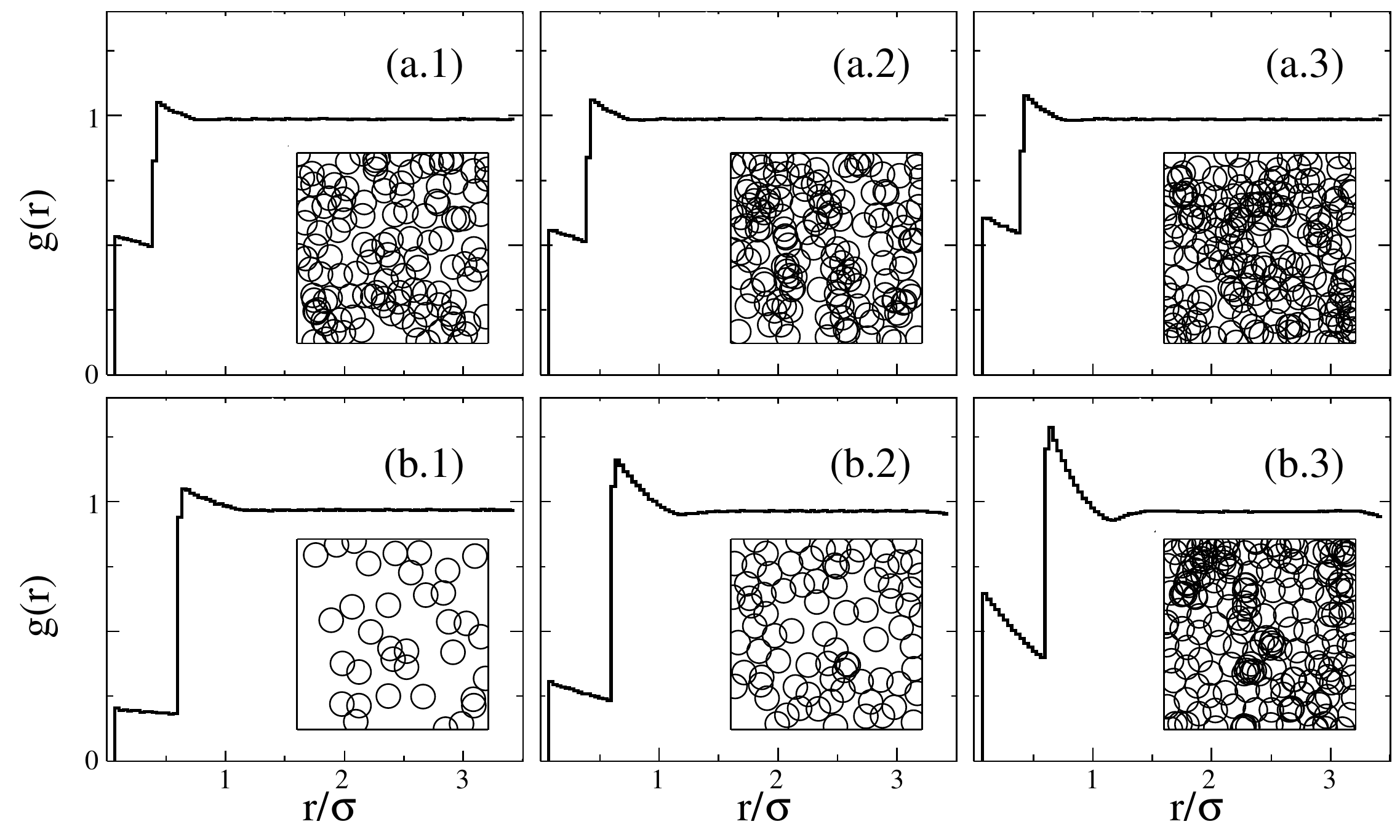}
\caption{Samples of systems for the CSCSPM along with their corresponding radial distribution function. Note that in (b.2) and (b.3), we observe a transition from nonideal gas to liquid-like structures.}
\label{fig:cscspm-gr}
\end{figure}

In Fig.~\ref{fig:cscspm-gr}, we show samples of core-shell-color string systems generated by using the aforementioned algorithm together with their corresponding radial distribution function, which is obtained as the average over $10^6$ realizations.

\begin{figure}[H]
\centering
\includegraphics[scale=0.35]{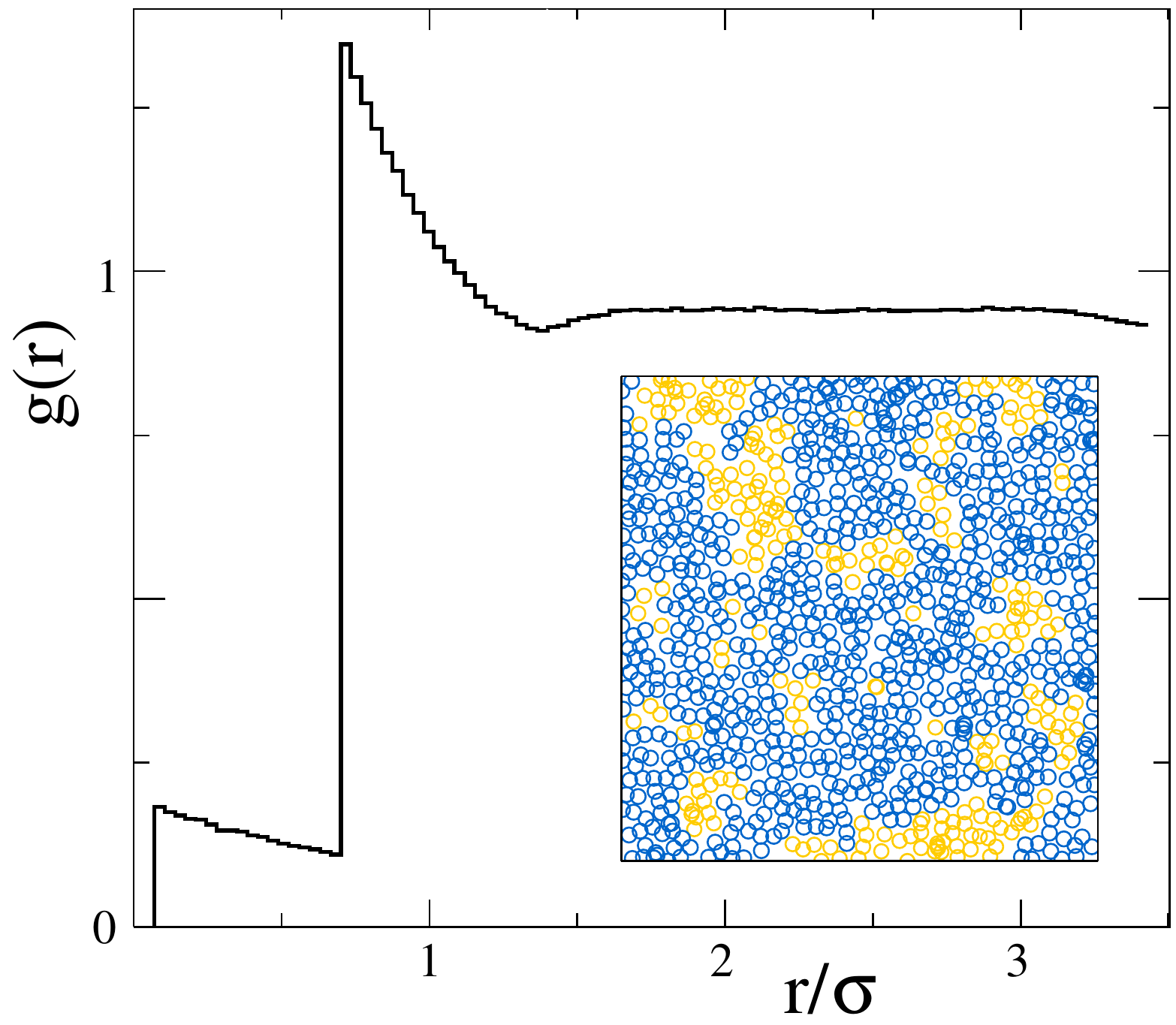}
\caption{Sample of a system that shows a liquid-like structure together with the spanning cluster. This occurs with a difference in transition temperatures (configurational and geometric) lower than 1 MeV.}
\label{fig:cscspm-sample}
\end{figure}

It is remarkable that the CSCSPM allows structures more complex than the ideal gas. In ref. \cite{JBP2021} was found that exists values of $\lambda$ and $q_\lambda$ at which the systems have liquid-like structures, as shown in Fig.~\ref{fig:cscspm-gr} (b.3).
It is worth mentioning that the CSCSPM predicts values of $\lambda$ and $q_\lambda$ at which the structural and geometrical transitions occur almost simultaneously, with a difference of temperatures of 1 MeV.
Figure~\ref{fig:cscspm-sample} shows a sample of a CSCSPM system exhibiting a liquid-like structure together with the emergence of the spanning cluster. 

\subsection{Color glass condensate}

\begin{figure}[H]
\centering
\includegraphics[scale=0.5]{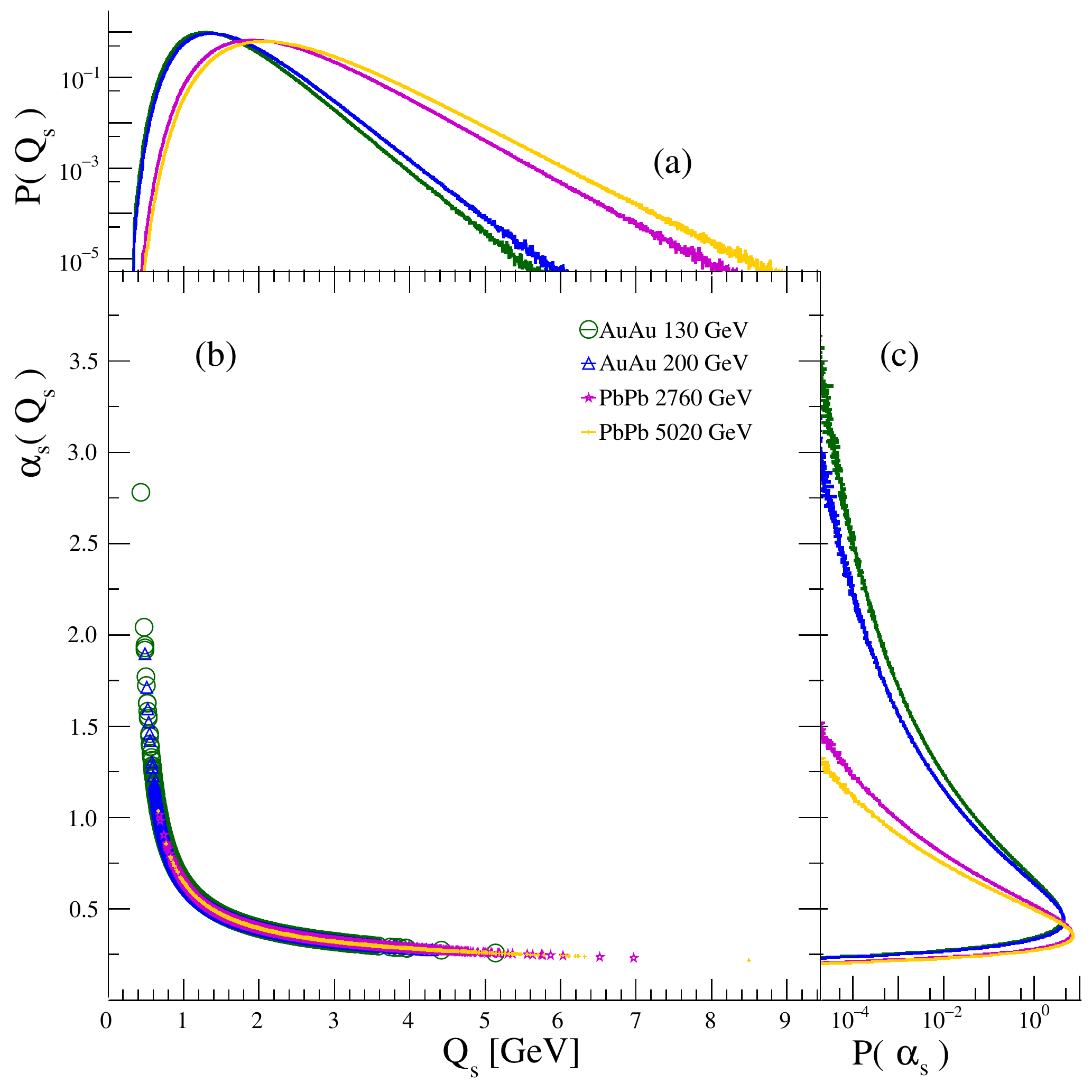}
\caption{(a) Distribution of $Q_S$. (b) Sample of $\alpha_s$-values as a function of $Q_s$. (c) Probability distribution  function of $\alpha_s$ of the systems under study: AuAu at $\sqrt{s} =$ 130 GeV (green lines and circles), AuAu at $\sqrt{s} =$ 200 GeV (blue lines and triangles), PbPb $\sqrt{s} =$ 2760 GeV (magenta lines and stars), and PbPb $\sqrt{s} =$ 5020 GeV (yellow lines and crosses.). }
\label{fig:alphaQs}
\end{figure}

To analyze the structure of systems given by the CGC model, it is required to know the distribution of the saturation scale $Q_s^2$, and the running coupling $\alpha_s$, which provide the initial conditions of the distribution of gluons and their interaction. In this work, we consider that $Q_s^2$ fluctuates according to the Log-normal distribution given by Eq.~\eqref{eq.PQs}. Then, the values of $Qs^2$ are simulated via the normal distribution, that is,
$Q_s^2=\langle Q_s^2 \rangle e^x$, where $x$ is a random number taken from a normal distribution $\mathcal{N}(0,1)$, and $\langle Q_s^2 \rangle$ is the average value of $Q_s^2$. The determination of $\langle Q_s^2 \rangle$ involves analyzing experimental data.
We use the modified minimal subtraction scheme ($\overline{MS}$) to compute the values of $\alpha_s$ at zero order. Thus
\begin{equation}
\alpha_{s}(Q_s^2) = \frac{4\pi}{\beta_0 \ln(Q_s^2 / \Lambda_\text{QCD}^2)},    
\label{eq:alpha}
\end{equation}
with $\beta_0=11-2n_f/3$, and $\Lambda_\text{QCD}^2$ being the QCD scale.
For the estimation of $\beta_0$ we use the value of the flavor number $n_f=3$, and $\Lambda_\text{QCD}=0.332$ GeV \cite{ParticleDataGroup:2016lqr}.
Notice that $\Lambda_\text{QCD}$ works as a cutoff for the values of $Q_s$ since lower values of $Q_s$ produce nonphysical values of $\alpha_s$.
In Fig.~\ref{fig:alphaQs}, we show a sample of $Q_s$ values and the corresponding value of $\alpha_s$ computed using \eqref{eq:alpha}, together with their histograms.

\begin{figure}[H]
\centering
\includegraphics[scale=0.25]{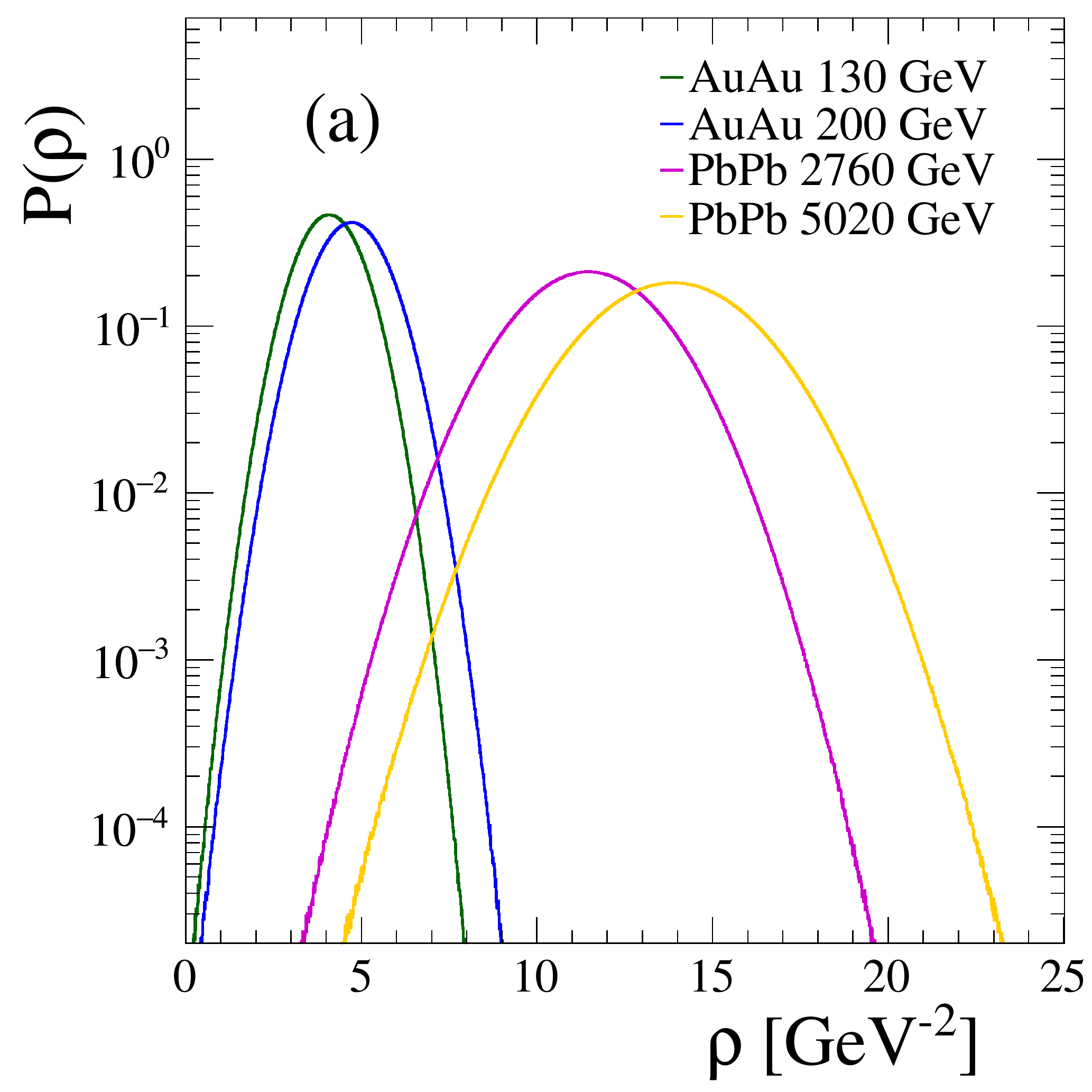}
\includegraphics[scale=0.25]{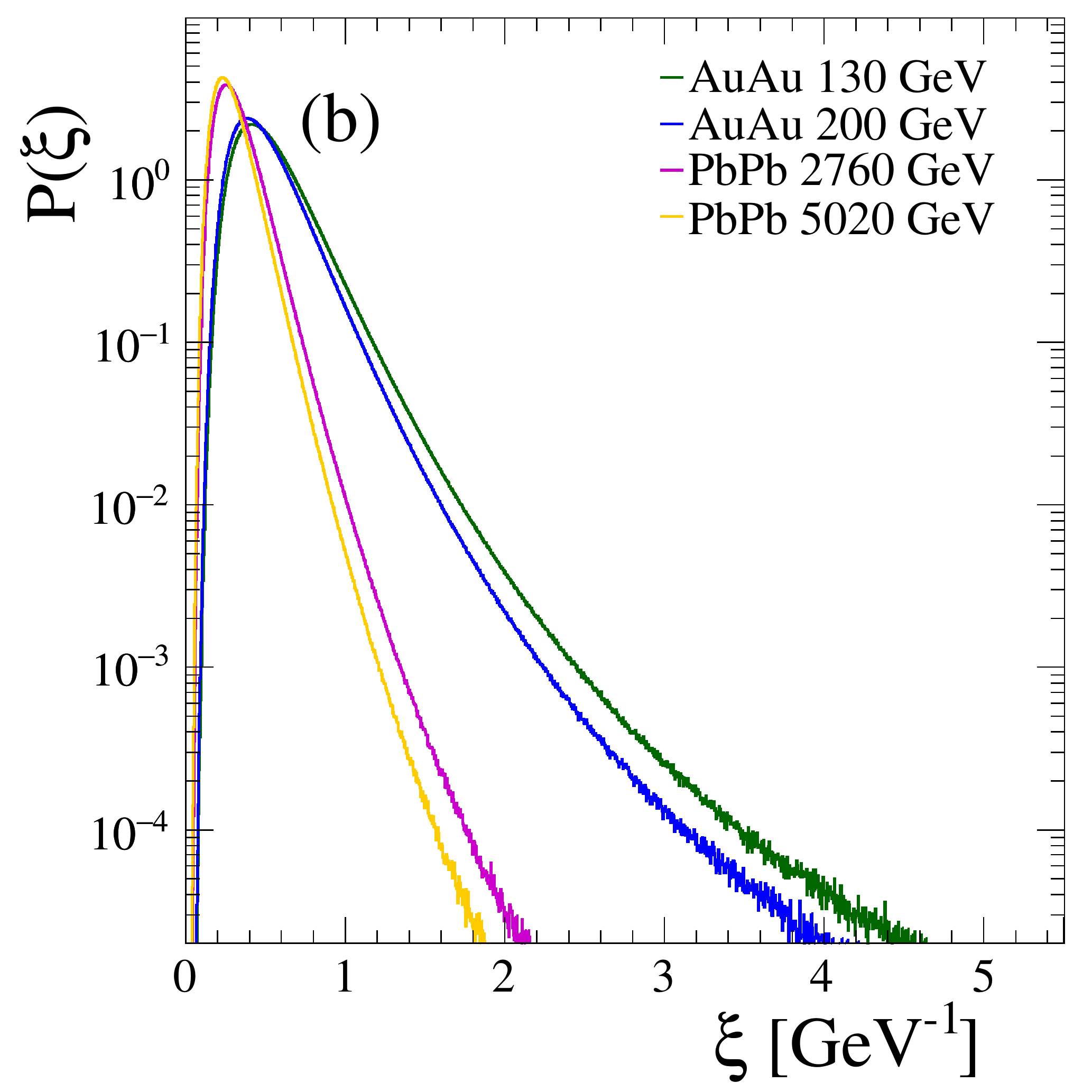}
\caption{Discrete histograms for simulations. (a) the minimum distances between gluon centers and (b) the gluon density of the system. }
\label{fig:gluondensity}
\end{figure}

In a nuclear collision, the Lorentz-contracted nuclei lead the partons to be confined to a flat region on the transverse plane. The occupied Glasma fields generated after the collisions are assumed to be proportional to the number of particles produced in a central pseudorapidity region at the saturation scale \cite{Kharzeev:2004if}, which for low transverse momentum values is given by $S Q_s^2 / \alpha_s$ \cite{Kharzeev:2000ph,Schenke:2013dpa}, where $S$ is the overlapping area of nuclei. Taking these relations into account, the gluon density 
is 
\begin{equation}
\rho=\frac{Q_s^2}{\alpha_s}.
\end{equation}
Due to the saturation scale and coupling running are random variables, the gluon number density is also expected to exhibit random fluctuations, as depicted in Figure \ref{fig:gluondensity} (a).
In consequence, the average minimum distance between gluons (computed from the gluon cross section) $\xi=1/\sqrt{\rho}$ is also a random variable, as shown in Figure \ref{fig:gluondensity} (b).

\begin{figure}[H]
\centering
\includegraphics[scale=0.25]{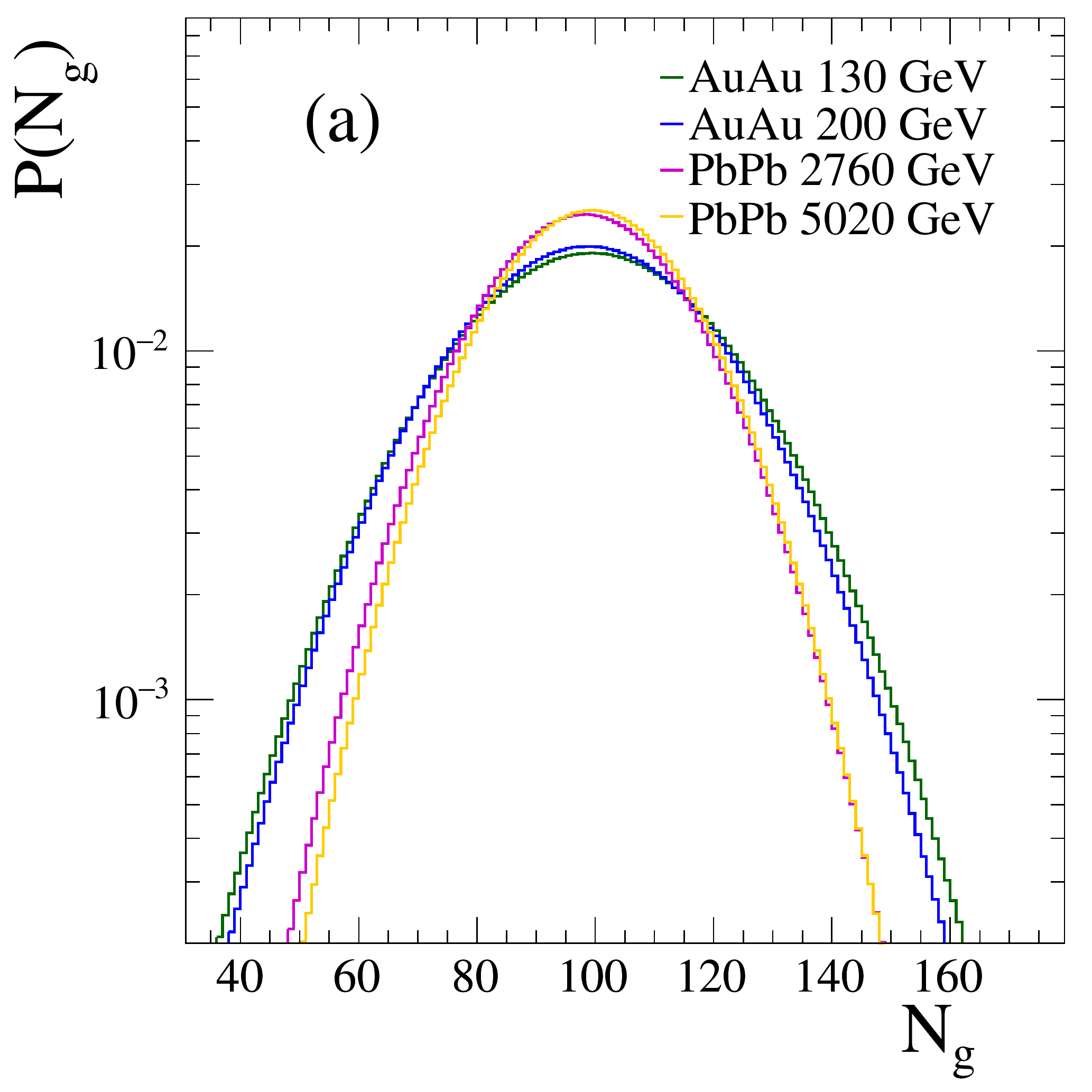}
\includegraphics[scale=0.25]{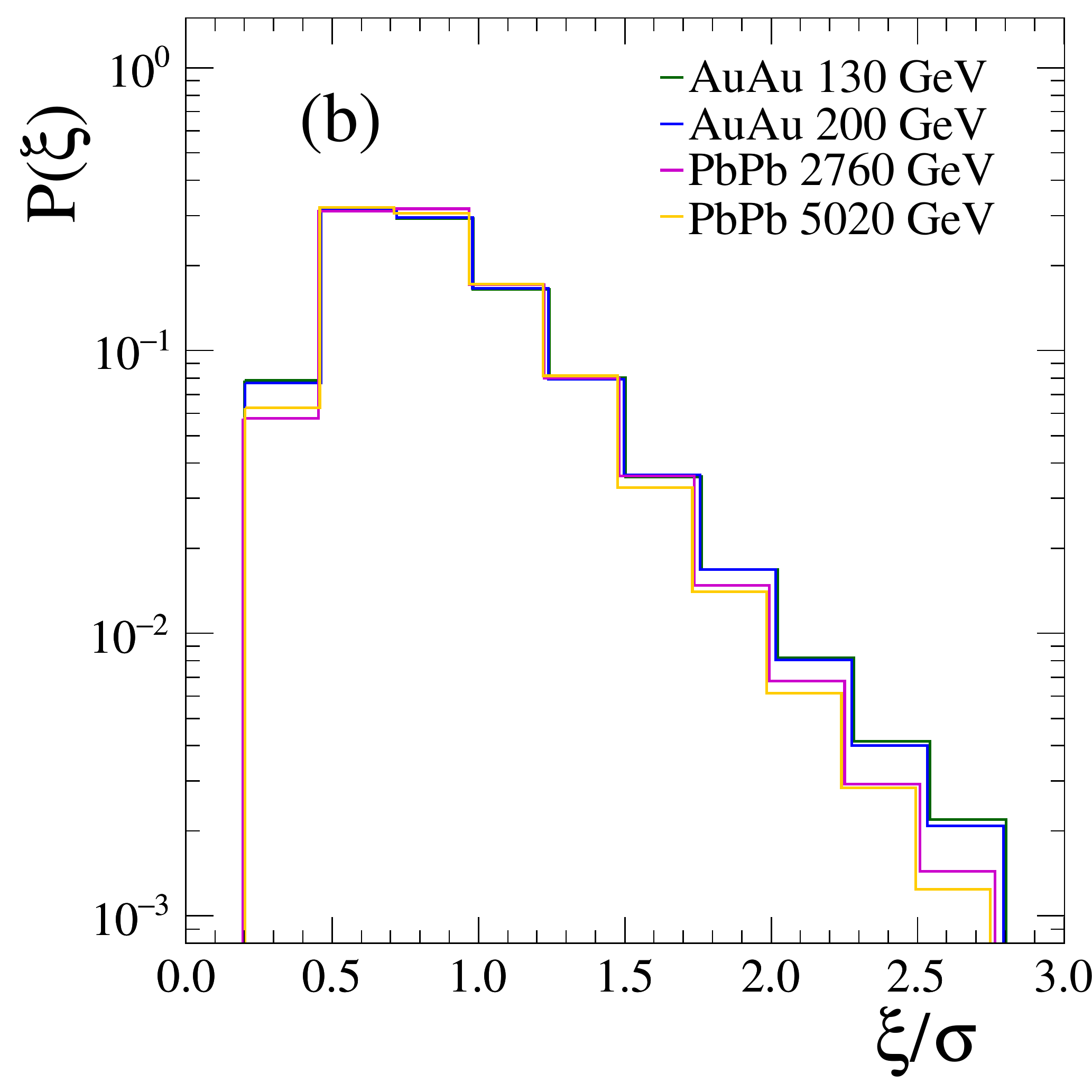}
\caption{Used histograms for simulation (a) the number of gluons in the square and (b) classes of minimum distances among gluons normalized by the average diameter of a gluon.} 
\label{fig:dist_simulations}
\end{figure}

For our simulations, we assume that the system is homogeneous and isotropic for AA central collisions (0-6\% centrality) and we use as a length scale the average value of the gluon diameter, $\sigma=2\langle r_g \rangle=2\langle 1/\sqrt{\pi \rho} \rangle$.
Following the same procedure as in the CSPM and the CSCSPM, we construct a square box of length side $L=8\sigma$. Thus, the number of gluons in the system is $N_g=64\sigma^2 \rho$.
Moreover, the probability of observing a system with exactly $N_g$ gluons is calculated as usual
\begin{equation}
P(N_g)=\int_{X} P(\rho) d\rho
\end{equation}
with $X$ being the interval of $\rho$-values satisfying $\lfloor L^2 \rho \rfloor=N_g$.
Figure~\ref{fig:dist_simulations} (a) shows the distribution mass probability for the gluon number obtained from our simulations. This distribution takes into account all the fluctuations arising from the conditions of the CGC model and is relevant for computing the average of the radial distribution function.

To ensure valid configurations of gluon positions in our simulations, we use a criterion based on the probability density function of the minimal distance between nearest neighbor gluons. Similarly to the number gluon $N_g$, we discretize that probability density function as follows. We consider the interval of $\xi$-values satisfying $P(\xi)>0.01$. For the systems analyzed in this manuscript, this event has a probability of occurrence of 0.9968, 0.9971, 0.9985, and 0.9986, respectively. Then, we divide the interval into ten equally sized sub-intervals, where the probability of observing the $i$-subdivision is given by
\begin{equation}
P(i)=\int_{X_i} P(\xi) d\xi \left / \int_{\bigcup_{i=1}^{10} X_i} P(\xi) d \xi \right. \,,
\end{equation}
which produces the histograms depicted in Fig.~\ref{fig:dist_simulations} (b).
This approach has two restrictions on the random variable $\xi$ with zero probability, namely, (i) $\xi/\sigma <0.2$, and (ii) $\xi/\sigma >2.8$.

\begin{figure}[H]
\centering
\includegraphics[scale=0.68]{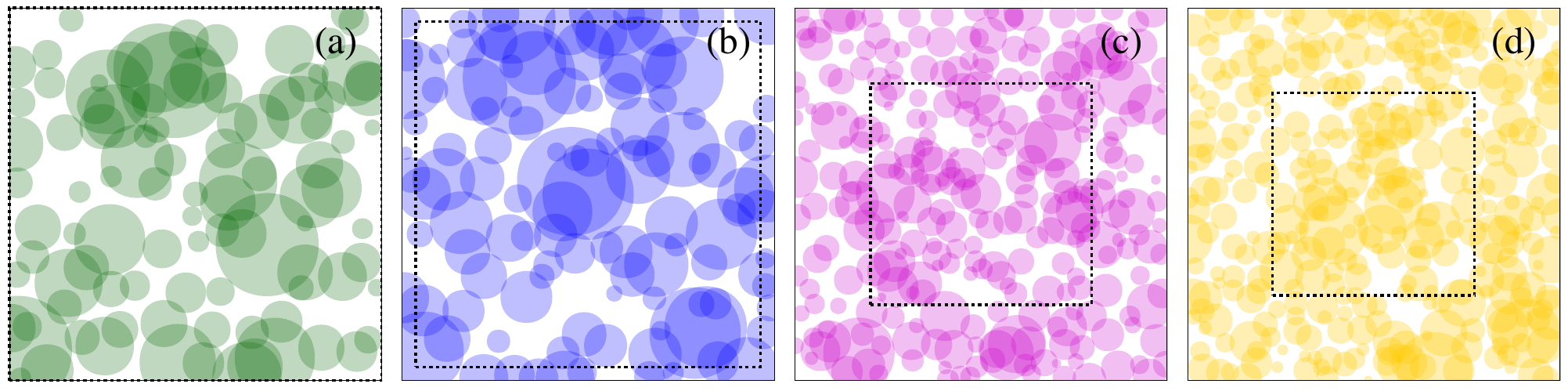}
\caption{Samples of generated systems in the picture of the CGC after the thermalization for (a) AuAu collisions at $\sqrt{s}=$130 GeV, (b) AuAu collisions at $\sqrt{s}=$200 GeV, (c) PbPb collisions at $\sqrt{s}=$2760 GeV, and (d) PbPb collisions at $\sqrt{s}=$5020 GeV. These systems correspond to a square box of around 1fm$^2$. The inner dashed square is the size of the simulated system for the purposes of the computation of the radial distribution function.}
\label{fig:samplesCGC}
\end{figure}

We initialize the simulation by allocating a trial disk at the center of a square box of side $L=8\sigma=16\langle r_g \rangle$.
Then, we add $N-1$ gluons  one by one ensuring that the minimal distance between nearest neighbor gluons satisfies $0.2<\xi/\sigma<2.8$.
After the addition of the $N$ gluons, we \textit{thermalize} the systems by moving an arbitrary gluon (except the trial gluon). 
For this test gluons, we generate a virtual random position on a neighborhood of the actual position and consider periodic boundary conditions. Thus, we calculate the distance $\xi'$ to the nearest gluon. Next, we generate a uniformly random number on the interval $(0,1)$. The virtual position is accepted if $P(\xi')>x$, with $P(\xi')$ taken from the distribution on Fig.~\ref{fig:dist_simulations} (b). Otherwise, the position is rejected, and we repeat this methodology for another gluon. This procedure is repeated $10^4$ moves per particle.
After thermalization, we measure the number of particles $n(r, N)$ within a distance between $r$ and $\Delta r$ from the trial gluon with $N$ particles in the system.
In Fig.~\ref{fig:samplesCGC}, we show samples of the systems after thermalization for AuAu and PbPb collisions at $\sqrt{s}=$130, 200 GeV and $\sqrt{s}=$2760, 5020 GeV, respectively.
For consistency, we use the same parameters as those used for the CSPM and CSCSPM models to construct the plot of $g(r)$.
We generate new valid configurations by moving a random gluon as described before and measure $n(r, N)$ again after 100 moves per gluon.
Finally, the average of the number of gluons $n(r, N)$ after $10^4$ simulation runs.

\begin{figure}[H]
\centering
\includegraphics[scale=0.5]{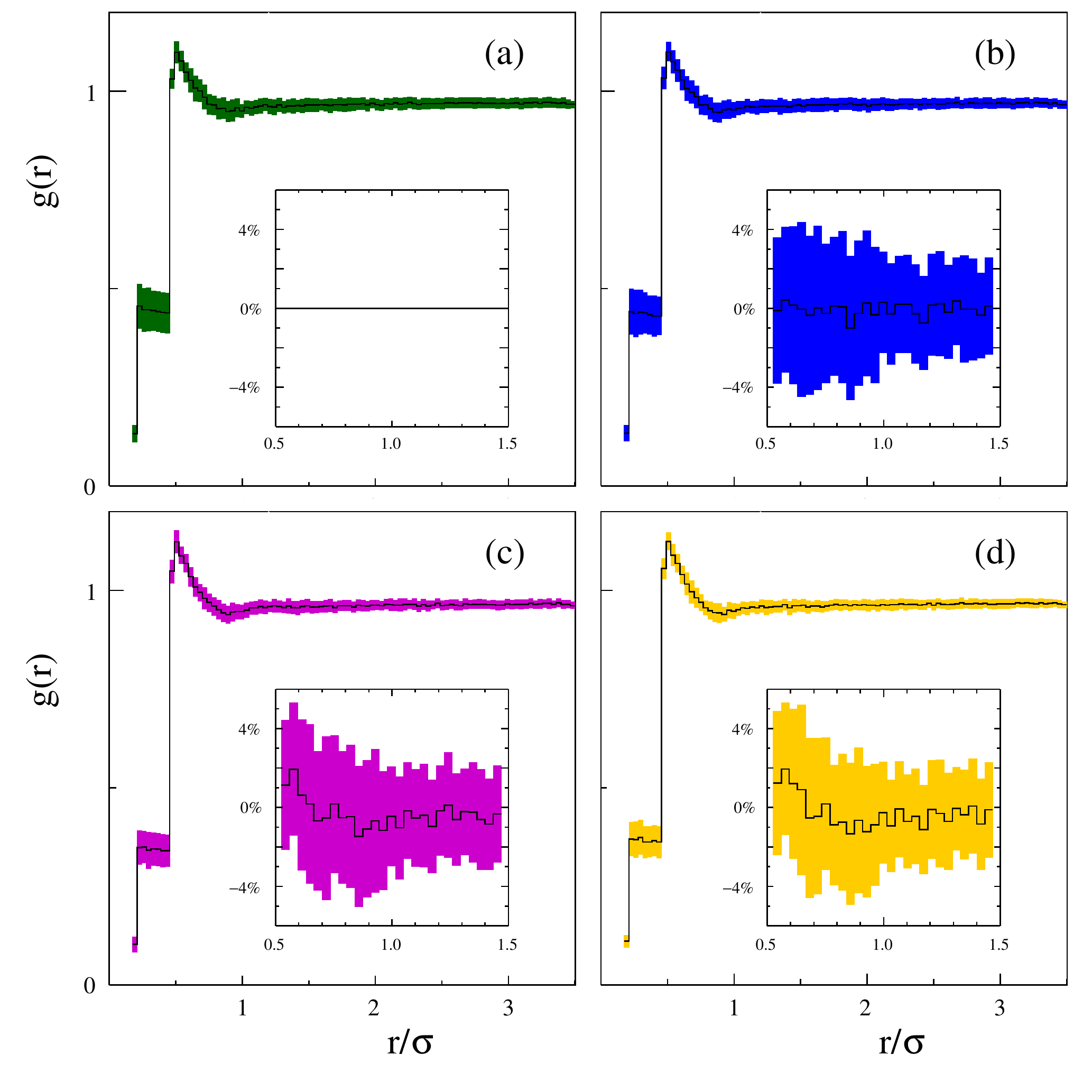}
\caption{Radial distribution function of systems in the picture of the CGC for (a) AuAu collisions at $\sqrt{s}=$130 GeV, (b) AuAu collisions at $\sqrt{s}=$200 GeV, (c) PbPb collisions at $\sqrt{s}=$2760 GeV, and (d) PbPb collisions at $\sqrt{s}=$5020 GeV. The inner plots correspond to the percentage of deviation of the radial distribution function from the one obtained for AuAu collisions at $\sqrt{s}=$130 GeV.}
    \label{fig:CGCstructure}
\end{figure}

We recall that the number of gluons on the systems fluctuates for particular conditions of center of mass energy, centrality classification, pseudorapidity, and nucleon number, among others.
To account for these fluctuations, we determine the average of the radial distribution function as follows
\begin{equation}
g(r)= \sum_{N=N_\text{min}}^{N_\text{max}} \frac{L^2 n(r, N)}{N(2\pi \Delta r (r+0.5\Delta r)+\pi \Delta r^2)} P(N)
\left /  \sum_{N=N_\text{min}}^{N_\text{max}} P(N) \right.,
\end{equation}
where $P(N)$ is the corresponding distribution of the number of gluons in Fig.~\ref{fig:dist_simulations} (a), $N_\text{min}=\lfloor \langle N_g \rangle \rfloor - 3 \left \lfloor \sqrt{\text{var}(N_g)} \right \rfloor$, and $N_\text{max}=\lfloor \langle N_g \rangle \rfloor + 3 \left \lfloor \sqrt{\text{var}(N_g)} \right \rfloor$.
Figure~\ref{fig:CGCstructure} summarizes our results of the radial distribution function for AuAu and PbPb central collisions at $\sqrt{s} =$ 130, 200 GeV and $\sqrt{s} =$ 2760, 5020 GeV, respectively

\section{Discussion and conclusions}\label{conclu}

In this paper, we have determined the structure of the medium formed in heavy ion collisions under the picture of the color string percolation model, the core-shell-color string percolation model, and the color glass condensate framework.
We did this by analyzing the radial distribution function of the particular configuration of the distributed transverse representation of color flux tubes on each model: color strings in the CSPM and CSCSPM and gluons in the CGC.

It was found that the color string percolation model is structured as an ideal gas because of the fully overlapping condition of the string.
By introducing a repulsive-like interaction between strings, the CSCSPM produces more complex structures, resembling a gas of interacting particles or a liquid-like fluid. The CSCSPM implements two new characteristics of the color strings: a concentric core region and a probability that a color string accepts core-core interaction.
 This implies a classification of the strings as soft and hard, which act like ideal particles or hard disks with diameter $\lambda \sigma$, respectively.
A fine-tuning of the CSCSPM parameters $(\lambda, q_\lambda)$ allows the systems to exhibit a structural phase transition at the same time as a geometrical phase transition. The former implies a transition from the gas structure to the liquid-like one, while the latter means the well accepted formation of the spanning cluster of color string, which can be interpreted as the onset of the quark-gluon plasma formation.
The CSCSPM predicts conditions that both transitions occur almost simultaneously, with a difference of temperature of 1 MeV.

Other systems analyzed were those pictured by the color glass condensate.
Here, we used the stochastic CGC framework that considers fluctuations on the saturation scale $Q_s^2$.
This also produces fluctuations in the running coupling, and then in all the characteristics of the CGC, such as the density, the size of gluons, and minimal distances between nearest neighbor gluons, among others.
We found that the distribution of gluons for AuAu central collisions at RHIC energies (130 and 200 GeV) corresponds to a gas of interacting particles.
On the other hand, for PbPb central collisions at LHC energies (2760 and 5020 GeV), we found evidence that the systems adopt a liquid-like structure. However, more simulations are needed to smooth the radial distribution function in Fig.~\ref{fig:CGCstructure} and apply a similar analysis on the $g(r)$ as described in Ref.~\cite{JBP2021}.
It is worth noting that while we simulated homogeneous and isotropic system representations at the center of the transverse plane to the collision, the system may dilute at the edges, resulting in a radial distribution function that resembles that of a diluted interacting gas and eventually an ideal gas (flat function).

%
%
\section*{Acknowledgments}

This work was partially supported by the grants PID2020-116567GB-C22 and CEX2019-000904-S funded by
MCIN/AEI/10.13039/501100011033. Bogar D\'iaz acknowledges support from the CONEX-Plus programme funded by Universidad Carlos III de Madrid and the European Union's Horizon 2020 research and innovation programme under the Marie Sklodowska-Curie Grant Agreement No. 801538. 
This work was partially funded by Consejo Nacional  de Ciencia y Tecnología (CONACyT-México) under the project CF-2019/2042,
graduated fellowships grant numbers 645654 and 1140160, and postdoctoral fellowship grant number 289198. We thank Carlos Pajares for his valuable comments and fruitful discussions.

%
%

%
%

\bibliographystyle{JHEP}
\bibliography{MCS}

\end{document}